# Spin-transfer switching and thermal stability in an FePt/Au/FePt nanopillar prepared by alternate monatomic layer deposition

Kay Yakushiji, Shinji Yuasa, Taro Nagahama, Akio Fukushima,

Hitoshi Kubota, Toshikazu Katayama and Koji Ando

*National Institute of Advanced Industrial Science and Technology (AIST), Nanoelectronics Research Institute, 1-1-1 Umezono, Tsukuba, 305-8568 Japan*

Abstract:

We fabricated a current-perpendicular-to-plane giant magnetoresistance (CPP-GMR) nanopillar with a 1 nm-thick FePt free layer having perpendicular anisotropy using the alternate monatomic layer deposition method. Nanopillars consisting of [Fe(1 monolayer(ML))/Pt(1 ML)]$_n$ (*n*: the number of the alternation period) ferromagnetic layers and an Au spacer layer showed spin-transfer induced switching at room temperature. An average critical switching current density ($J_{c0}$) of 1.1 x 10$^7$ A/cm$^2$ with a large thermal stability parameter ($\Delta$) of 60 was obtained in a nanopillar with a free-layer thickness of 1.02 nm (*n* = 3) and a pillar diameter of 110 nm. The ultrathin free-layer with high perpendicular anisotropy is effective to obtain both large $\Delta$ and small $J_c$.

Corresponding author:
K. Yakushiji
E-mail address: k-yakushiji@aist.go.jp

Magnetization switching induced by a spin-polarized current[1,2] has attracted a great deal of attention because it can be used as a scalable writing technology for high-density spin-torque switching-type magnetoresistive random access memory (so-called Spin-RAM). So far, the spin-transfer switching has been intensively studied experimentally using magnetoresistive devices with in-plane magnetized materials such as CoFe or CoFeB[3-9]. Two important properties predicting a higher density Spin-RAM application are low critical current density ($J_c$) of spin-torque switching and thermal stability parameter ($\Delta$) of the memory cell. The $\Delta$ above 60 is required for the data retention time of the stored information to extend beyond 10 years, a benchmark that should be satisfied in a non-volatile memory device. A critical current density ($J_c$) at the order of $10^6$ A/cm$^2$ is required in Gbit-scale high-density Spin-RAM because of the necessity of shrinking the size of the path transistors serially connected to the magnetoresistive cells. However, these two requirements have never been satisfied simultaneously because the $\Delta$ and the $J_c$ exist in a trade-off relation. To satisfy these two requirements, we have focused our study on a magnetoresistive device with perpendicular magnetic anisotropy. L1$_0$-ordered alloys are known to exhibit large perpendicular anisotropy ($K_u$), which realizes larger thermal stability ($\Delta = K_u V / k_B T$) than those of ordinary in-plane magnetized materials. Furthermore, a magnetoresistive nanopillar with perpendicular anisotropy has the possibility of showing a small $J_c$ because the demagnetizing field is collinear to the perpendicular anisotropy field. Although basic properties of spin-torque switching in nanopillars with perpendicular anisotropy have been studied in recent years[10-15], there have been few systematic studies to clarify the detailed mechanisms.

We have been interested in an ultrathin L1$_0$-ordered FePt layer prepared by the monatomic alternate layer deposition method. An ideally ordered FePt possesses $\Delta$ above 60 even in a few nanometer-sized particles due to the largest class of $K_u$ which exceeds 5x10$^7$ (erg/cc). The deposition method is effective in preparing an ultrathin film (< 2nm) without deteriorating the film quality[16-19]. During the deposition in an ultrahigh vacuum, an active surface migration occurs even at a relatively low substrate temperature around 200 ºC, which gives rise to the formation of an epitaxially grown ordered film with an atomically flat surface. In this study, we employed a monatomic alternate layer deposition method to prepare high-quality ultrathin FePt switching layers with perpendicular magnetization, which are expected to realize both large $\Delta$ and small $J_c$. We investigated spin-transfer switching and thermal stability in epitaxial [Fe(1 monolayer(ML))/Pt(1 ML)]$_{12}$ / Au / [Fe(1 ML)/Pt(1 ML)]$_n$ current-perpendicular-to-plane giant magnetoresistance (CPP-GMR) nanopillars.

Films were deposited by the molecular beam epitaxy (MBE) method. A Cr/Au buffer was deposited at 250 ºC. Then Fe of 1 ML, a [Fe(1 ML)/Pt(1 ML)]$_n$ pinned layer with $n$ = 12 (4.0 nm- thick), Fe of 2 ML, a 2.7 nm- thick Au spacer layer, Fe of 1ML, and a [Fe(1 ML)/Pt(1 ML)]$_n$ free layer were successively deposited on the buffer at 200 ºC. The deposition of [Fe(1 ML)/Pt(1 ML)]$_n$ started with Fe and ended with the Pt layer. The number of alternation periods $n$ for the free layer was varied at 3, 4, 5 and 6: the corresponding thicknesses were from 1.02 ($n$ = 3) to 2.04 nm ($n$ = 6). These varied stacks were prepared in a single substrate by shutter control during deposition. Finally, Ru was deposited as a capping layer by sputtering. Microfabrication to form CPP-GMR devices was performed by the combination of electron-beam lithography, optical lithography, rf-sputtering and Ar ion milling. Pillars were prepared in a circular shape with various diameters from 100 nm to 170 nm. In addition to CPP-GMR stacks, [Fe(1ML)/Pt(1ML)]$_n$/Fe bilayer films were also prepared under the same deposition conditions in order to study the magnetic property of the films.

Figure 1 shows polar Kerr hysteresis loops of the [Fe(1ML)/Pt(1ML)]$_8$/Fe($t_{Fe}$) bilayer film deposited on a single substrate at 200 ºC. The thickness of Fe, $t_{Fe}$, was varied from 0.43nm (3 ML) to 1.29nm (9 ML) while that of FePt, $t_{FePt}$, was fixed at 2.7 nm ($n$ = 8). A bilayer with $t_{Fe}$ = 3 ML shows a sharp transition with good squareness in the magnetization loop. The coercivity ($H_c$) of the films decreases with increasing $t_{Fe}$, and the bilayer with $t_{Fe}$ = 9 ML lost the squareness in its loop. This result indicates that the addition of the Fe layer onto the FePt layer reduces the net perpendicular anisotropy of the bilayer, and the thickness ratio of FePt ($t_{FePt}$ = 2.7 nm) and Fe ($t_{Fe}$ = 9 ML was considered to be around a critical point toward in-plane anisotropy. We prepared other combinations of bilayers (not shown in the figure): $t_{FePt}$ = 2.04 nm ($n$ = 6) and 1.36 nm ($n$ = 4) with $t_{Fe}$ varying from 0.43nm (3 ML) to 1.29nm (9 ML). In all these combinations, $H_c$ decreased with an increasing thickness ratio: $t_{Fe}/t_{FePt}$. The critical points of net anisotropy for $t_{FePt}$ = 2.04 nm and a 1.36 nm bilayer were $t_{Fe}$ = 1.00 nm (7 ML) and 0.72 nm (5ML), respectively. The results indicate that the net anisotropy and resulting net $H_c$ of a bilayer are tunable by the thickness ratio of $t_{Fe}/t_{FePt}$.

In the CPP-GMR stack, the bottom Fe(1 ML) / [Fe(1 ML)/Pt(1 ML)]$_{12}$ / Fe(2 ML)

and top Fe(1 ML) / [Fe(1 ML)/Pt(1 ML)]$_n$ layers corresponded to pinned and free layers, respectively. A cross section of a CPP-GMR cell is illustrated in fig. 2(a). We confirmed from the reflection high-energy electron diffraction (RHEED) patterns that Fe and Pt mono-layers were epitaxially grown in an alternating pattern during the [Fe(1ML)/Pt(1ML)]$_n$ preparation. A resistance – magnetic field (*R-H*) minor loop at RT for a cell with a free-layer thickness ($t_{\text{free}}$) of 1.36 nm ($n = 4$) and a pillar diameter of 100 nm is shown in fig. 2 (b). The center of the loop shifts about 1100 Oe because of the dipole interaction between the free and pinned layers. It obtains an MR ratio of 1.1 % and an resistance-area product (RA) of 0.06 Ωμm$^2$. Domain motion is considered to govern the magnetization reversal of the free layer resulting in the sharp transitions in the *R-H* loop. Coherent rotation cannot occur in the present cell size because it should take place only in a pillar diameter below 50 nm[20].

Next, we observed the current-induced magnetization switching. An external field of 4000 Oe, which was large enough to obtain an anti-parallel (AP) state, was applied. Then the field was decreased to 1100 Oe (the center of *R-H* loop), and resistance-current (*R-I*) measurement was carried out. The initial direction of a current pulse of 100-ms duration was positive; that is, electrons went from the pinned to the free layer. Spin-torque switching for both anti-parallel to parallel (AP to P) and parallel to anti-parallel (P to AP) states was clearly observed with sharp transitions as shown in fig. 2 (c). Fig. 2 (d) represents the current pulse duration (*t*) dependence of the switching current ($I_c$). The critical current ($I_{c0}$) and Δ were estimated from the extrapolated $I_c$ value at $t = 1$ ns and the slopes of the relationship[21,22] to be $I_{c0}^{\text{AP to P}} = 1.60$ mA, $I_{c0}^{\text{P to AP}} = -1.45$ mA, and $\Delta^{\text{AP to P}} = 60$, $\Delta^{\text{P to AP}} = 75$, respectively. $J_{c0}^{\text{AP to P}}$ and $J_{c0}^{\text{P to AP}}$ were estimated from $I_{c0}$ and the pillar diameter (100nm) to be 2.0 x 10$^7$ and -1.8 x 10$^7$ A/cm$^2$, respectively.

Figure 3 plots the average critical current density $J_{c0}^{\text{av}}$ ($\equiv \left( J_{C0}^{\text{AP to P}} - J_{C0}^{\text{P to AP}} \right)/2$) versus average thermal stability $\Delta^{\text{av}}$ ($\equiv \left( \Delta^{\text{AP to P}} + \Delta^{\text{P to AP}} \right)/2$) for various cells. Although the data were scattered, nearly all the cells showed small $J_{c0}^{\text{av}}$ while maintaining practically large $\Delta^{\text{av}}$ above 60. In a cell with $t_{\text{free}} = 1.02$ nm and a pillar diameter of 110 nm, $J_{c0}^{\text{av}} = 1.1$ x 10$^7$ A/cm$^2$ and $\Delta^{\text{av}} = 60$ were realized. We consider that the high perpendicular anisotropy of FePt gave rise to a practical Δ even in a cell with the thinnest free layer ($t_{\text{free}} = 1.02$ nm), and use of such an ultrathin free layer led to a small $J_{c0}$ in spite of the small MR and large damping property of FePt. The scattered data might relate to the complicated switching mechanism of the free layer, which is not a simple coherent rotation, but a spin-transfer-induced domain wall motion.

In conclusion, we fabricated epitaxial [Fe(1ML)/Pt(1ML)]$_{12}$ / Au / [Fe(1ML)/Pt(1ML)]$_n$ ($n$ = 3, 4, 5 and 6)-based CPP-GMR nanopillars with perpendicular magnetic anisotropy and investigated their spin-transfer switching behaviors. We succeeded in preparing 1.02 nm-thick ($n$ = 3) perpendicular magnetized FePt film at a deposition temperature of 200 ºC. CPP-GMR nanopillars with an ultrathin free layer showed effective spin-transfer switching: $J_{c0}^{av}$ = 1.1 x 10$^7$ A/cm$^2$ and $\Delta^{av}$ = 60 were realized in a cell with $t_{free}$ = 1.02 nm ($n$ = 3) and a pillar size of 110 nm. Consequently the very thin free layer with the high perpendicular anisotropy led to the large $\Delta$ and small $J_{c0}$.

This work was supported by New Energy and Industrial Technology Development Organization (NEDO). We thank S. Mitani (Tohoku Univ.) for giving an advice on film preparation and Y. Suzuki (Osaka Univ.) and A. Deac (NIST) for useful discussions.

Figure captions

Fig. 1

Polar Kerr hysteresis loops for [Fe(1 monolayer(ML)) / Pt(1ML)]$_8$ / Fe($t_{Fe}$) bilayers. A magnetic field was applied perpendicular to the film plane. The thickness of the Fe($t_{Fe}$) was varied, at 3, 5, 7 or 9 ML.

Fig. 2

(a) Schematic illustration of a cell of FePt-based CPP-GMR. The thickness of the [Fe(1ML) / Pt(1ML)]$_n$ free layer ($t_{free}$) was varied from 1.02 to 2.04 nm in a single substrate. (b) Typical results for the magnetic field ($H$) dependence of resistance ($R$) at RT for a cell with $t_{free}$ = 1.36 nm ($n$ = 4) and a pillar diameter of 100 nm. (c) The result of spin-transfer switching (*R-I*) at $H$ = 1100 Oe with a current pulse duration of 100 ms. (d) Current pulse duration ($t$) dependence of switching current ($I_c$) of AP to P (upper) and P to AP (lower). The lines represent the guides assuming $I_c \propto \log t$ based on the experimental plots.

Fig. 3

Average critical current density ($J_{c0}^{av}$) versus average thermal stability ($\Delta^{av}$) for various cells.

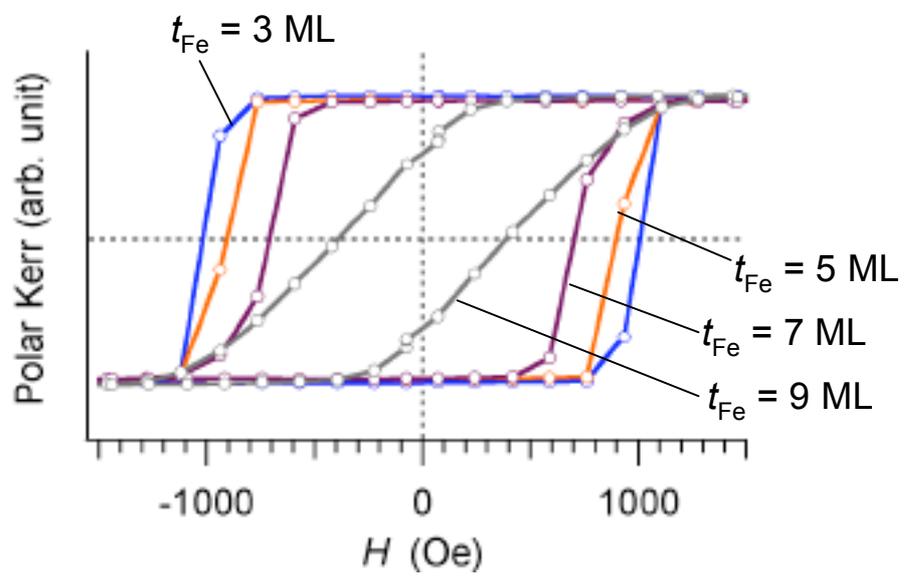

Fig. 1   YAKUSHIJI

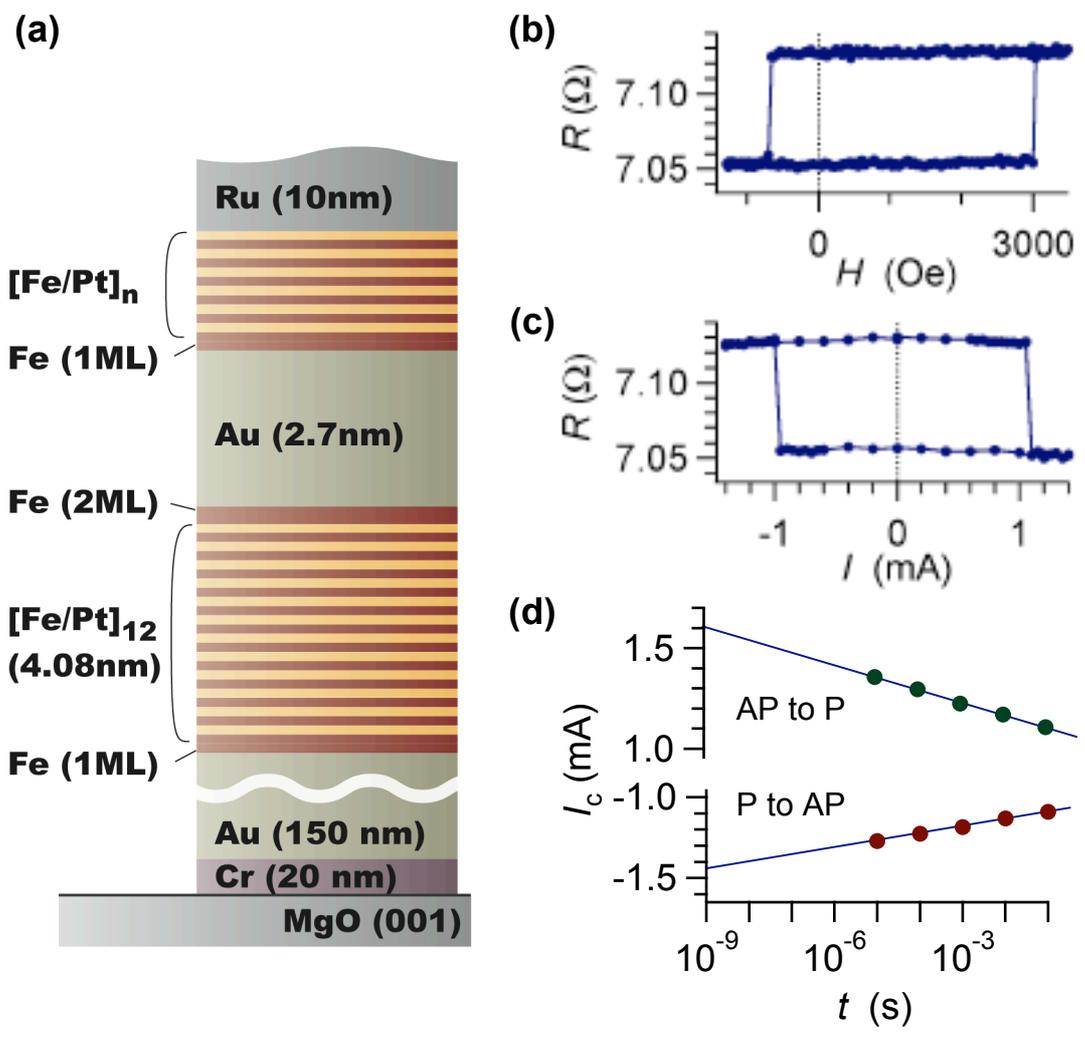

Fig. 2   YAKUSHIJI

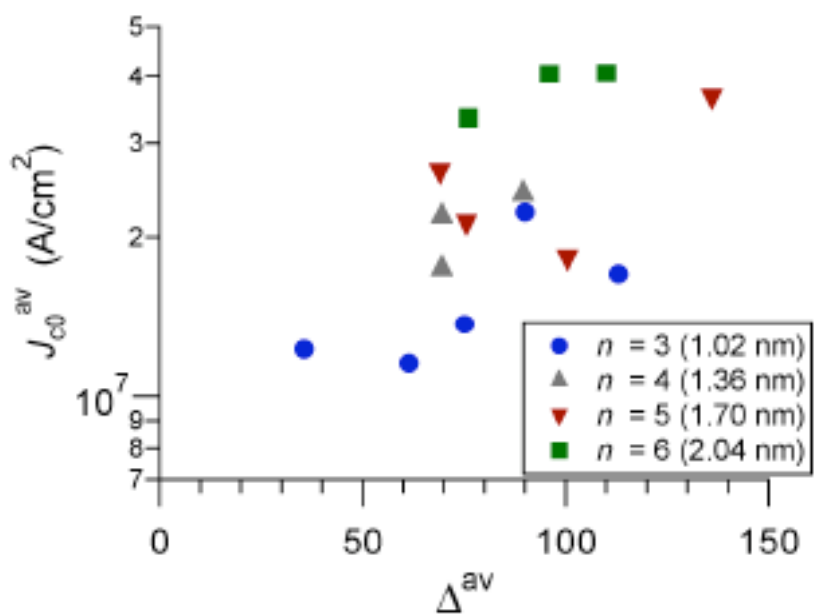

Fig. 3   YAKUSHIJI